\documentclass[prd, preprint, showpacs, showkeys, nofootinbib]{revtex4}    % (no)showpacs:= (no) display PACS number (default = no)
\usepackage{amssymb}
\usepackage{amsmath}
\usepackage{graphicx}
\usepackage[dvips]{epsfig}
\usepackage{hyperref}

\newcommand{\eq}[1]{Eq.~(\ref{eq:#1})}

\newcommand{\fig}[1]{Fig.~(\ref{fig:#1})}
\newcommand{\scn}[1]{Sec.~(\ref{sec:#1})}

\newcommand{\diby}[2]{\ensuremath{\frac{\partial #1}{\partial #2}}}
\newcommand{\equa}[1]{\begin{equation} #1 \end{equation}}

\def\lf {\ensuremath{\left}}
\def\rt {\ensuremath{\right}}
\def\ra {\ensuremath{\rightarrow}}

\def\qand {\ensuremath{\quad\text{and}}}

\newcommand{\qii}[2]{\ensuremath{q^{#1}_{#2}}}
\newcommand{\bqii}[2]{\ensuremath{\bar{q}^{#1}_{#2}}}
\newcommand{\dbqii}[2]{\ensuremath{\dot{\bar{q}}^{#1}_{#2}}}
\newcommand{\taii}[3]{\ensuremath{\lf. t_{#1} \rt.^{#2}_{#3}}}
\newcommand{\pmw} [1] {\ensuremath{\partial_{#1} \omega }}

\def\bphi {\ensuremath{\bar{\phi}}}
\def\vphi {\ensuremath{\mathbf{\Phi}}}
\def\phid {\ensuremath{\phi^\dag}}
\def\vphid {\ensuremath{\mathbf{\Phi}^\dag}}
\def\vphig {\ensuremath{\mathbf{\Phi}_g}}
\def\vphigd {\ensuremath{\mathbf{\Phi}_g^\dag}}
\def\vG {\ensuremath{\mathbf{G}}}
\def\vGd {\ensuremath{\mathbf{G}^\dag}}
\def\vA {\ensuremath{\mathbf{A}}}
\def\vAd {\ensuremath{\mathbf{A}^\dag}}
\def\vt {\ensuremath{\mathbf{t}}}
\def\vtd {\ensuremath{\mathbf{t}^\dag}}

\begin{document}

% -------------------------------- The Front Matter -------------------------------------------
\title{Implementing Mach's Principle Using Gauge Theory}
\date{\today}
\author{Sean Gryb}
\affiliation{Perimeter Institute for Theoretical Physics\\Waterloo, Ontario N2L 2Y5, Canada}
\affiliation{Department of Physics and Astronomy, University of Waterloo\\Waterloo, Ontario N2L 3G1, Canada}
\email{sgryb@perimeterinstitute.ca}
\pacs{04.20.Cv}
\keywords{Mach's Principle; Barbour-Bertotti Theory; gauge theory; relationalism, quantum gravity}

% --------------------------------------- Abstract ---------------------------------------------
\begin{abstract}
    We reformulate an approach fist given by Barbour and Bertotti (BB) for implementing Mach's principle for nonrelativistic particles. This reformulation can deal with arbitrary symmetry groups and finite group elements. Applying these techniques to U(1) and SU(N) invariant scalar field theories, we show that BB's proposal is nearly equivalent to defining a covariant derivative using a dynamical connection. We then propose a modified version of the BB method which implements Mach's principle using gauge theory techniques and argue that this modified method is equivalent to the original. Given this connection between the particle models and Yang-Mills theories, we consider the effect of dynamic curvature as a possible generalization of the BB scheme. Since the BB method can be used as a novel way of deriving geometrodynamics, the connection with gauge theory may shed new light on the gauge properties of the gravitational field.
\end{abstract}

\maketitle
\tableofcontents
% --------------------------------- End of Front Matter ----------------------------------------

% ==================================== Body of Paper ===========================================

\section{Introduction}

Mach's criticism of the Newtonian picture of spacetime in Chap. 2, Sec. 6 of the \emph{Mechanics} \cite{mach:mechanics} has had a profound influence on the development of theoretical physics. Despite the beauty, simplicity, and power of his arguments, Mach failed to provide a concrete theory or precise prescription for implementing his ideas. Perhaps for this reason, finding consistent statements of what is now called Mach's Principle is a difficult task. Of the many physicists and philosophers who have been profoundly influenced by the idea, one of the most notable is Einstein. As is well known, Einstein was inspired by the idea that the laws of physics should only depend on observable things and used this Machian idea as the foundation of General Relativity. Though guided by impressive intuition, Einstein did not have access to, nor did he correctly formulate, a precise implementation of Mach's principle. One modern proposal, offered by Barbour and Bertotti \cite{barbourbertotti:mach,barbour_el_al:rel_wo_rel,barbour_el_al:physical_dof,barbour:eot}, for such an implementation leads directly to General Relativity  when it is applied to a theory of dynamical geometry.\footnote{In $\mathbb{R}\times\Sigma$ topology with $\partial\Sigma = 0$.} This proposal, called \emph{best-matching}, is a general framework designed to remove nonphysical absolute background spatial structures from dynamical theories in a way that is meant to carefully implement Mach's principle. To my knowledge, best-matching is the most careful and systematic proposal for implementing Mach's principle. In this work, we study the mathematical structure of best-matching by developing a general framework and by testing this framework on enlightening examples. By comparing best-matching to Yang-Mills gauge theories, we find that best-matching is equivalent to defining a dynamical connection on configuration space. Thus, it is a framework for implementing Mach's principle using the techniques of gauge theory\footnote{For a slightly different view of Yang-Mills theory that uses a Jacobi-type action see \cite{anderson:gauge_theory} or, for Yang-Mills coupled to GR, see \cite{anderson:rel_wo_rel_vec}.}.

Best-matching owes its name to the fact that it is a procedure whereby two arbitrary configurations of the universe, shifted relative to each other, are compared by ``matching'' them, in a least squares sense, by shifting them in the direction of spatial symmetries. The method takes the absolute coordinates of some objects in the universe and shifts them to their best-matched position where they are referred to as \emph{corrected coordinates} \cite{barbour:scale_inv_particles}. Unfortunately, the original \emph{corrected coordinate method} (CCM) suffers from a lack of generality in the sense that only specific types of coordinate symmetries are considered and the framework is restricted to infinitesimal symmetry transformations. The latter deficiency results in difficulties arising from the appearance of awkward terms in the infinitesimal expansion. In this paper, we introduce a new formulation of the CCM\footnote{This \emph{reformulation} should not be confused with the \emph{modified} CCM that we introduce later.} that can deal easily with the awkward terms and also allows for a direct comparison to gauge theory.

One can understand physically why we should expect a link between best-matching and gauge theory. The method provides a prescription for computing the ``true'' infinitesimal difference between two configurations of the universe even though these configurations may be translated arbitrarily along the direction of some symmetry. This is a prescription for computing a derivative on a configuration space which is foliated by equivalence classes generated by some symmetry group, $\mathcal{G}$. If one thinks of these equivalence classes as fibers over a base manifold then best-matching is way of defining a connection on a $\mathcal{G}$-bundle. In particular, we will find that the least squares-type matching will lead us directly to a dynamical, but pure gauge, flat connection. Thus, best-matching is a way of gauging a spacetime symmetry by choosing a flat connection\footnote{See \cite{gomes:gauge_gravity} for further views on this.}. This realization will lead us to two considerations: 1) Can best-matching be generalized by choosing a dynamical connection like, for example, the Yang-Mills connection? and 2) We will see that in order to make the relationship between the CCM and gauge theory more natural it is necessary to modify the CCM slightly without changing the physical theory. This modification is simply a removal of mathematically awkward terms that appear in the original method. Hence, is this \emph{modified CCM} a more appropriate mathematical realization of best-matching?

The original CCM was developed by Barbour and collaborators in \cite{barbour:scale_inv_particles} and \cite{barbour_el_al:scale_inv_gravity} as a way of constructing a spatially relational theory in accordance with Mach's principle. To achieve temporal Relationalism, Jacobi's principle is used. However, it was found in \cite{sg:mach_time} that the CCM can also be used, in a mathematically equivalent way, to construct a temporally relational theory. This suggests that the CCM may be useful in a much broader context to implement a more general form of ``Relationalism''. We explore this possibility in the current paper by looking at several examples of theories with different symmetries. In particular, we study field theories with internal symmetries such as complex scalar fields invariant under the fundamental representations of U(1) and SU(N). We find that the CCM makes the internal phases arbitrary while letting the norm of the fields evolve according to the Klein-Gordon equation. These examples are enlightening for the following reasons: 1) by comparing the CCM to Yang-Mills gauge theories we can deduce a precise relationship between the variables used in the original CCM and the components of a dynamical connection, 2) the analogy with Yang-Mills suggests possible ways to generalize best-matching to include dynamical curvature, and 3) the connection with gauge theory allows us to import intuition and mathematical tools from gauge theory to best-matching where they can be used to study the gravitational field.

\section{The Corrected Coordinate Method and Best-Matching}

We would now like to describe our formulation of the CCM and briefly compare it to older formulation of the technique used in \cite{barbour:scale_inv_particles} to check that the essential elements are the same. We will make a distinction between the \emph{corrected coordinate method}, which is a general method for achieving spatial relationalism, and \emph{best-matching}, which is a general technique that uses the CCM but also uses Jacobi's principle to implement temporal relationalism. We will make this distinction because we wish to use the CCM in a broader context to be able to: a) deal with internal symmetries and b) deal with cases where the temporal relationalism is \emph{not} implemented using Jacobi's principle.

\subsection{Corrected Coordinates in the Particle Model}\label{sec:cc_particle}

Consider a system of $N$ particles in $3$ dimensional Euclidean space. Label the instantaneous positions of these particles by $q^i_I(\lambda)$ where lower case Roman indices range from 1 to 3 and label spatial coordinates while upper case Roman indices range from 1 to $N$ and label distinct particles. The collection of all $q^i_I(\lambda_0)$ labels a point in configuration space at a specific value of the parameter $\lambda=\lambda_0$. Consider two such configuration space points $q^i_I(\lambda_1)$ and $q^i_I(\lambda_2)$.
\begin{figure}\includegraphics[width=.75\linewidth]{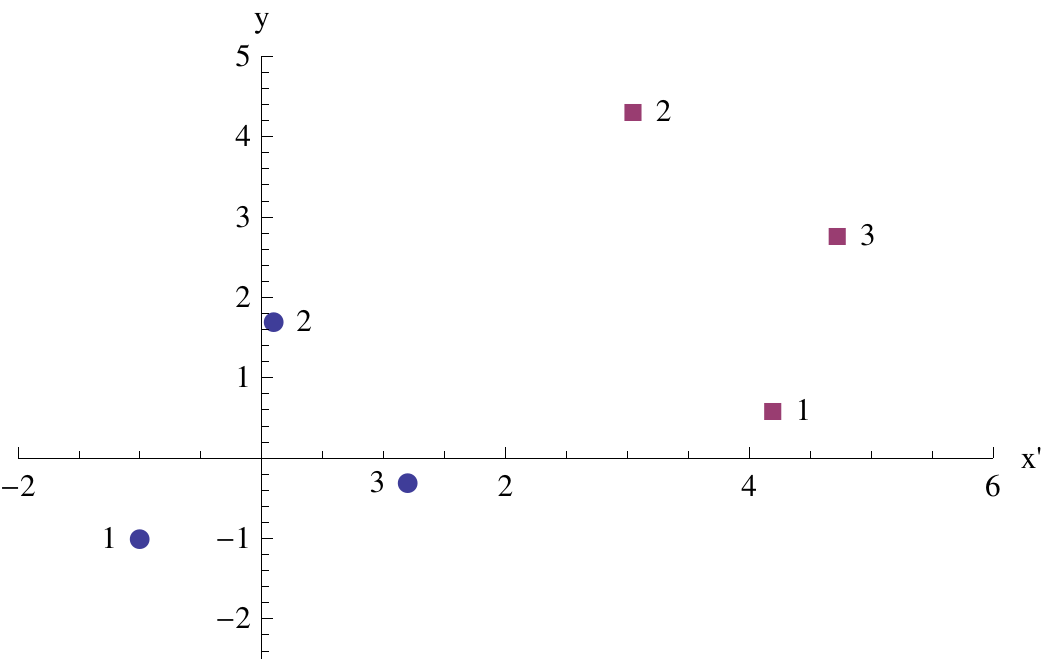}\\
  \caption{The \emph{circles} represent the $q^i_I(\lambda_0)$. They are the positions of different particles at some value of the parameter $\lambda =\lambda_0$. The \emph{squares} are the positions of the same particles at some later value of $\lambda =\lambda_1$. Note that rotations and translations don't change the ratios of the lengths of the triangle.}
  \label{fig:one}
\end{figure}
\fig{one} shows an example of what these configurations might look like in the case of only three particles. According to Barbour and Bertotti's interpretation of Mach's principle \cite{barbourbertotti:mach}, a sensible physical theory should only depend on quantities that are directly observable to the particles in the system. In this case, these quantities should be the ratio's of the distances $r^i_{IJ}$ between the $I^{\text{th}}$ and $J^{\text{th}}$ particles. However, attempts by Hofmann, Reisser, Schrodinger, and others\footnote{For English translations, details on original publications, and useful editorial comments on these early attempts, see \cite{barbour:newton_2_mach}.} to formulate theories directly in terms of the $r^i_{IJ}$ have proven to be problematic as they lead to mass anisotropy far above experimental limits. Instead, we wish to formulate the theory in terms of the nonphysical Euclidean coordinates $q^i_I(\lambda)$ but impose constraints to reduce the unphysical \emph{Absolute Configuration Space} degrees of freedom, consisting of the Euclidean positions, to the more physical \emph{Relational Configuration Space} variables, made of the physically accessible ratios or the $r^i_{IJ}$\footnote{See \cite{gergely:geometry_BB1, gergely:geometry_BB2, anderson:found_part_dyn} for a detailed description of the CS's.}.

Before describing Barbour and Bertotti's method for accomplishing this reduction we should clarify the reason for its necessity. After all, if our theory admits global symmetries then we know from Noether's theorem that there will be conserved quantities. We are then free to invent auxiliary fields that generate these symmetries and whose conjugate momenta are equal to these conserved quantities. In this form, the auxiliary fields are, in the language of Lanczos \cite{lanczos:mechanics}, ignorable coordinates and there exists a well known procedure due to Routh for eliminating them. Unfortunately, the Routh procedure can only be applied when the Poisson brackets between the ignorable coordinates vanish. That is, Routhian reduction is only valid when the symmetry group is Abelian. For further details on this see, for example Sec. 4.3 of \cite{abraham_marsden:mechanics}. In the present context, we are interested in a more powerful method that can deal with general Lie groups. Best-matching provides such a framework.

To accomplish the reduction, we notice that the $q^i_I$'s can be arbitrarily rotated and translated without affecting the $r^i_{IJ}$. If $G^i_j$ is an element of the Euclidean group, consisting of 3 dimensional rotations and translations, and $\omega^\alpha$ are $\lambda$-dependent group parameters (with Greek indices ranging from 1 to the dimension of the group), we can apply the active transformation
\equa{\label{eq:barq}
   \bqii{i}{I}(\lambda) =  G^i_j(\omega^\alpha(\lambda)) \qii{j}{I}(\lambda),
}
without changing the physically observable quantities available to the system. Because it is only the \emph{ratios} of the $r^i_{IJ}$ that are observable, we could add scale transformations to the list of symmetries that do not affect the physical system. In that case, we require that $G$ be an element of the similarity group. In terms of the local algebra generators $\lf. t_\alpha \rt.^i_j$, the $G$ can be written as:
\equa{\label{eq:Gij}
    G^i_j(\omega^\alpha(\lambda)) = \exp\lf\{{\omega^\alpha(\lambda) \lf. t_\alpha \rt.^i_j}\rt\}.
}
For the specific groups we are considering, the generators of the local algebra are given in Table~\ref{tbl:one}.
\begin{table}
    \centering
\begin{tabular}{| c | c | c |}
    \hline
    Group        & Dimension & $\lf. t_\alpha \rt.^i_j$ \\
    \hline
    translations &    3 $(\alpha = k = 1\hdots 3)$      & $\delta^i_j \partial_k $ \\
    \hline
    rotations    &    3 $(\alpha = k = 1\hdots 3)$     & $\epsilon_{ijm}q^m \partial_k$  \\
    \hline
    scale        &    1 $(\alpha = 0)$     & $\delta^i_j q^m \partial_m$ \\
    \hline
\end{tabular}\quad.
    \caption{The generators of the Similarity group.} \label{tbl:one}
\end{table}
The group parameters, $\omega^\alpha$, are called the \emph{auxiliary} fields. In Table~\ref{tbl:one}, we have suppressed particle labels since the generators act equally on every particle.

\begin{figure}\includegraphics[width=0.75\linewidth]{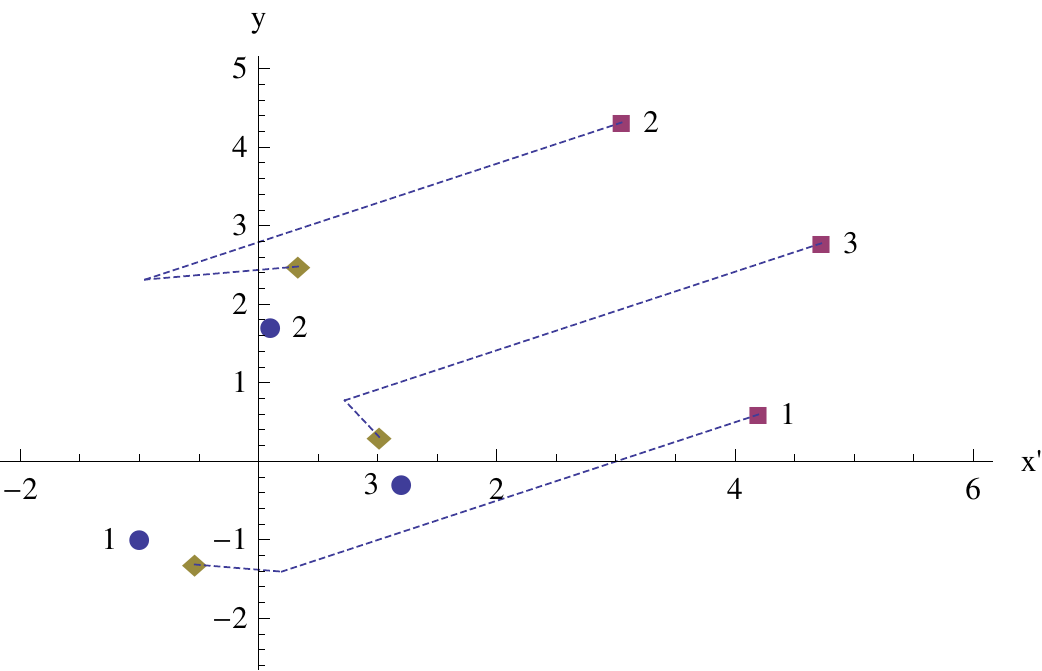}
  \caption{The \emph{diamonds} represent the $\bar{q}^i_I(\lambda_1)$. They are the best-matched positions of the $q^i_I(\lambda_1)$ (the \emph{squares}) after a series of translations and rotations. This puts them into a position where they can be democratically compared to the $q^i_I(\lambda_0)$ (the \emph{circles}).}
  \label{fig:two}
\end{figure}

With these definitions, an observer inside the system will not be able to distinguish between the $q$'s and the $\bar{q}$'s. Thus, when observers compare the two points $\qii{i}{I}(\lambda_1)$ and $\qii{i}{I}(\lambda_2)$ they could equally well, from their point of view, be comparing the points $\bqii{i}{I}(\lambda_1,\omega_1)$ and $\bqii{i}{I}(\lambda_1, \omega_2)$ for some arbitrary $\omega_1$ and $\omega_2$. \fig{two} shows an example of how the group parameters $\omega$ can be used to produce physically equivalent configurations.

We would like to have a procedure for determining the ``difference'' between two configurations at two different values of $\lambda$ (say $\lambda$ and $\lambda+\delta\lambda$) that does not depend on arbitrary symmetry transformations. A democratic solution is to choose the $\omega$'s such that the quantity
\equa{\label{eq:chi2}
    \chi^2 = \sum_I \lf( G_I^J(\lambda + \delta\lambda) q_J(\lambda+\delta\lambda) - G_I^J(\lambda) q_J(\lambda) \rt)^2
}
is minimized. In best-matching, $\chi^2$ is used in conjunction with Jacobi's principle to define a metric on configuration space. The CCM is a procedure that minimizes $\chi^2$ when $\delta\lambda$ is infinitesimal. Thus, the CCM implements a type of least-squares matching to points in the Absolute Configuration Space.

The way in which the CCM accomplishes this minimization is to modify the action principle of the absolute theory by everywhere substituting the normal coordinates $q^i_I$ with the \emph{corrected coordinates} $\bqii{i}{I} = G_I^J q^i_J$. This will modify the kinetic terms by sending the derivatives of $q$ with respect to $\lambda$ to:
\begin{align}
    \dot{q} \ra \dot{\bar q } &= \diby{(Gq)}{\lambda} \notag \\
                              &= G \dot{q} + \dot G q \notag \\
                              &= G (\dot q + \dot{\omega}^\alpha t_\alpha q).
\end{align}
where we have used matrix notation for multiplying the matrices $G$ and $t_\alpha$ with the column vector $q$ (note that spacetime indices have been suppressed). Because the $\omega$'s are arbitrary at the initial and final times, we require that a \emph{free endpoint} variation of the action with respect to the $\omega$'s vanish. In other words, we allow the variation of $\omega$ to be arbitrary on the boundary. If the action has a kinetic term, this will minimize the square of the quantity
\equa{\label{eq:min}
    G (\dot q + \dot{\omega}^\alpha t_\alpha q)
}
at all values of $\lambda$.

To see that this is equivalent to the least-squares minimization of (\ref{eq:chi2}), we can expand $G(\lambda + \delta\lambda)$ and $q(\lambda+\delta\lambda)$ in orders of $\delta\lambda$ and then collect the highest order terms. Noticing that all $\mathcal{O}(1)$ terms cancel and that $\dot{G} = G \dot{\omega}^\alpha t_\alpha$ according to the definition (\ref{eq:Gij}), we find that
\equa{
    \chi^2 = \sum_I \lf( G (\dot q + \dot{\omega}^\alpha t_\alpha q) \rt)^2 \delta\lambda^2 + \mathcal{O}(\delta\lambda^3).
}
Thus, minimizing $\chi^2$ is equivalent to minimizing (\ref{eq:min}) for infinitesimal $\delta\lambda$.

\begin{figure}\label{fig:three}
    \includegraphics[width=0.75\linewidth]{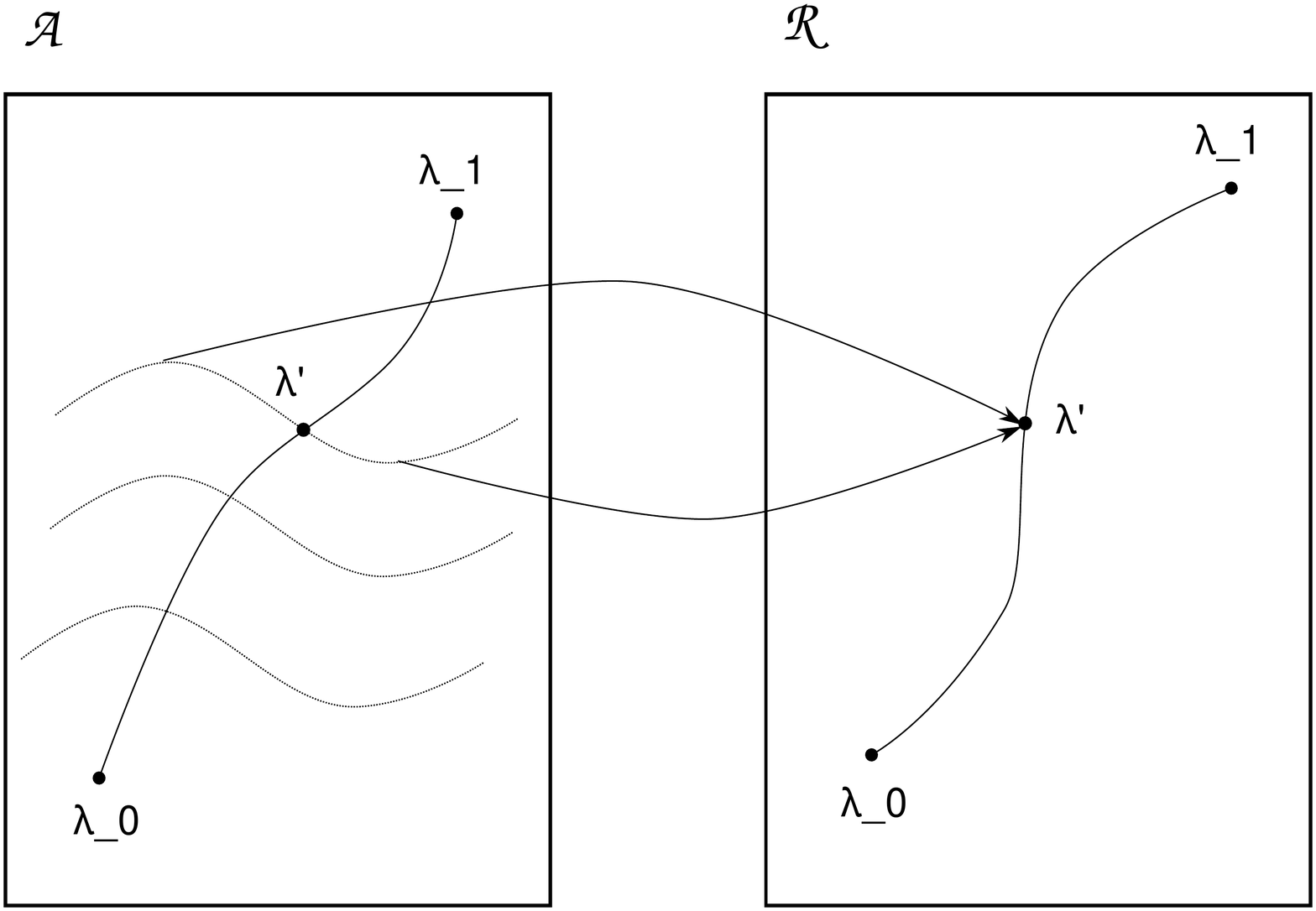}
    \caption{The Absolute Configuration Space, $\mathcal{A}$ is foliated by gauge orbits of equivalent configurations. These project down to a single point in the Relational Configuration Space, $\mathcal{R}$ where solutions of theory are represented by geodesics.}
\end{figure}

We are now in a position to identify a connection between the CCM and gauge theory. The theory is defined on an Absolute Configuration Space, $\mathcal{A}$, which is foliated by gauge orbits generated by a symmetry group $\mathcal{G}$. In our case, $\mathcal{G}$ is the similarity group. Each of these gauge orbits is an equivalence class of configurations of the system and can be projected down onto a single point on the Relational Configuration Space, $\mathcal{R}$. $\mathcal{R}$ can then be though of as the base manifold of a $\mathcal{G}$-bundle as is shown in \fig{three}. Solutions of the theory are paths formed by sections of the $\mathcal{G}$-bundle. They are geodesics when projected on $\mathcal{R}$ and represent the trajectory of the system between $\lambda_0$ and $\lambda_1$. The CCM is a procedure for defining a $\lambda$-derivative along such a path on the $\mathcal{G}$-bundle where $\lambda$ parameterizes the path. Thus, it is a definition of a connection along a path on the $\mathcal{G}$-bundle. We will see in \scn{U1sf} and \scn{SUNsf} that the choice of connection implied by the CCM is precisely that of a flat connection.

%The corrected coordinate method then consists of the following: it allows you to compare two configurations rotated by some nonphysical symmetry by introducing the corrected coordinates $\bar{q}(\lambda, \omega(\lambda))$ and freely varying $\omega(\lambda)$ until the difference $\lf|(\bqii{i}{I}(\lambda_2,\omega_2) - \bqii{i}{I}(\lambda_1,\omega_1))\rt|$ is minimized. In a dynamical theory, this is equivalent to substituting $\bar{q}$ for $q$ in an action with a suitable kinetic term (which will contain derivatives of $q$) and then finding the values of $q$ and $\omega$ whose variation leaves the action extremal. Because we have no reason to specify $\omega$ on the boundary of the integral, we must not require that it's variation on the boundary be zero. This means that we should perform a \emph{free endpoint variation} on $\omega$. This will be our general prescription for eliminating a nonphysical symmetry from a theory.

\subsection{Original Formulation of the CCM} \label{sec:org_form}

We will now compare our proposed formulation of the CCM to the original formulation to see that they are equivalent. In the original formulation of the CCM \cite{barbour:scale_inv_particles}, the corrected coordinates were defined as:
\equa{
    \bqii{i}{I} = q^{i}_{I} + \epsilon\lf( t^{i} + \epsilon^{ijk} r^{j} q^{k}_I + d q^i_I \rt)
}
where $\epsilon$ is infinitesimal and $\epsilon^{ijk}$ is the Levi-Civita symbol. $t^i$, $r^j$, and $d$ are $\lambda$-dependent parameters that determine, respectively, the amount of translation, rotation, and dilatation of the best-matched coordinates. This, of course, is completely equivalent to the definition (\ref{eq:barq}) with
\begin{align}
    \omega^\alpha &= \epsilon t^i,\quad\text{for }\alpha = 1\hdots 3, \notag \\
    \omega^\alpha &= \epsilon r^i,\quad\text{for }\alpha = 3\hdots 6, \text{and}\notag \\
    \omega^\alpha &= \epsilon d,\quad\text{for }\alpha = 7. \label{eq:new_2_old}
\end{align}

The downside of the original approach is that it deals only with infinitesimal transformations and is restricted to particular types of symmetry transformations. Our new formulation can handle large gauge transformations and general Lie groups which allows for a more general treatment. Furthermore, the derivatives of the corrected coordinates with respect to $\lambda$ are given by:
\equa{
    \dot{\bar q}^i_I = \dot{q}^{i}_{I} + \epsilon\lf( \dot{t}^{i} + \epsilon^{ijk} \dot{r}^{j} q^{k}_I + \dot{d} q^i_I \rt)+ \epsilon\lf( \epsilon^{ijk} r^{j} \dot{q}^{k}_I + d \dot{q}^i_I \rt).
}
The second term of order $\epsilon$, which contains $\dot{q}$, is awkward. It is difficult to deal with mathematically and seems to have no effect on the physical theory. Later we will propose a \emph{modified} version of the CCM which makes it clear why this term is not important but, for the moment, we notice that, in our new formulation, this term is hidden in the factor $G\dot q$ and is, thus, much easier to deal with mathematically.

\subsection{Barbour-Bertotti Theory}

We will now use the CCM to eliminate the Euclidean and scaling symmetries of classical mechanics. We will implement temporal Relationalism by invoking Jacobi's principle. This will be Barbour-Bertotti theory \cite{barbourbertotti:mach, barbour:scale_inv_particles} written in terms of our new formulation.

We start with a Jacobi action defined on what we have called our absolute configuration space:
\equa{
    S_J = \int_{\lambda_1}^{\lambda_2} d\lambda \, 2\,\sqrt{E - V(\qii{i}{I})}\sqrt{T(\dot{q}^i_I)},
}
where $T(\dot{q}^i_I) = \sum_I \frac{1}{2} m_I (\dot{q}^i_I)^2$ and a dot indicates differentiation with respect to the arbitrary parameter $\lambda$. As is well known \cite{barbourbertotti:mach, lanczos:mechanics}, this action will produce a temporally relational theory equivalent to standard Newtonian mechanics whose solutions are restricted to some total energy $E$\footnote{Note that, in the Barbour and Bertotti's approach to Jacobi's theory \cite{barbourbertotti:mach}, $E$ plays the role of a constant of Nature analogous to the role of the cosmological constant in GR. It is not determined by initial conditions as is the case in standard Newtonian theory.}. We now implement the CCM described in the previous section.

Our first step is to substitute the corrected coordinates $\bar{q}$'s for the $q$'s:
\equa{
    S_J = \int_{\lambda_1}^{\lambda_2} d\lambda \, 2\,\sqrt{E - V(\bqii{i}{I})}\sqrt{\sum_I \frac{1}{2} m_I (\dot{\bar{q}}^i_I)^2}.
}
A variation with respect to the $q$'s gives Newton's laws after making a gauge choice where the auxiliary fields $\omega^\alpha$ are equal to zero. This gauge choice is called the \emph{distinguished representation}. For more details on these equations of motion see \cite{barbour:scale_inv_particles} for the original formulation or \cite{sg:dirac_algebra} for the formulation proposed here.

We will now outline the \emph{free} endpoint variation with respect to the auxiliary fields $\omega^\alpha$ because it is not the usual variation used in physics and because it imposes the key relational ideas. Consider an arbitrary variation of $S_J$ with respect to $\omega^\alpha$:
\equa{
    \delta_\omega\, S_J = \int_{\lambda_0}^{\lambda_1} d\lambda \, \lf[ \frac{\partial \mathcal{L}_J}{\partial \omega^\alpha} -  \frac{d}{d\lambda}\lf(\frac{\partial \mathcal{L}_J}{\partial \dot{\omega}^\alpha}\rt) \rt]\, \delta\omega^\alpha + \lf[ \frac{\partial \mathcal{L}_J}{\partial \dot{\omega}^\alpha} \, \delta\omega^\alpha \rt]_{\lambda_0}^{\lambda_1} = 0.
}
The boundary term is usually set to zero by setting $\delta\omega^\alpha = 0$ on the boundary but, in this case, we have no reason to specify $\omega$ on the boundary leaving its variation there nonzero. In order to guarantee the vanishing of the variation of $S_J$, we can undo the integration by parts leaving us with
\equa{
	\delta_\omega\, S_J = \int_{\lambda_0}^{\lambda_1} d\lambda \, \lf[ \lf( \frac{\partial \mathcal{L}_J}{\partial \omega^\alpha} \rt)\, \delta\omega^\alpha + \lf(\frac{\partial \mathcal{L}_J}{\partial \dot{\omega}^\alpha}\rt)\delta\dot{\omega}^\alpha\rt] = 0.
}
For arbitrary variations, this is can be made to vanish only if
\begin{align}\label{eq:mach_cons}
	\frac{\partial \mathcal{L}_J}{\partial \dot{\omega}^\alpha} &= 0, \qand \\
	\diby{\mathcal{L}_J}{\omega^\alpha} &= 0
\end{align}
separately. This is equivalent to enforcing the usual Euler-Lagrange equations and the additional condition (\ref{eq:mach_cons}). (\ref{eq:mach_cons}) is the standard \emph{free endpoint condition} considered by \cite{barbourbertotti:mach, barbour:scale_inv_particles} and, because it imposes key Machian ideas, (\ref{eq:mach_cons}) has also been called the \emph{Mach condition} \cite{sg:mach_time}.

Evaluating the partial derivatives of the Mach condition gives
\equa{
    \sqrt{\frac{E-V}{T}} \sum_I m_I \dot{\bar{q}}^I_i \lf. t_\alpha \rt.^i_j \bar{q}^i_I = 0.
}
If we define $\pi^I_\alpha = m_I \bar{q}\prime^I_i \lf. t_\alpha \rt.^i_j \bar{q}^i_I$, with primes denoting differentiation with respect to the quantity $dt_e = d\lambda \sqrt{\frac{T}{E-V}}$\footnote{$dt_e$ is a natural, gauge invariant increment of time referred to as \emph{ephemeris time} in \cite{barbourbertotti:mach} and discussed in \cite{sg:emer_time_PI}.}, as the \emph{generalized momentum} of the $I^\text{th}$ particle then the Mach condition is just the vanishing of the \emph{total} generalized momentum. For translations,
\begin{align}
    \pi_\alpha^I &= m_I \bar{q}^I_i, &(\alpha &=1\hdots 3) \\
\intertext{is just the usual linear momentum. For the rotations,}
    \pi_\alpha^I &= m_I \epsilon_{ijk} \dot{\bar{q}}^j_I \bar{q}^k_I, &(\alpha &=4\hdots 6) \\
\intertext{is just the usual angular momentum. For the scale transformations,}
    \pi_\alpha^I &= m_I \dot{\bar{q}}_i^I \bar{q}^i_I = \frac{1}{2}\dot{I}_I, &(\alpha &=7)
\end{align}
is just Barbour's dilatational momentum \cite{barbour:scale_inv_particles} or the $\lambda$-derivative of the moment of inertia $I$ of the I$^{\text{th}}$ particle. Hence, the Mach condition imposes the vanishing of the total linear, angular, and dilatational momentum of the system, perfectly implementing Barbour and Bertotti's Machian program.

After imposing the Mach constrain, we must still impose the usual Euler-Lagrange equations. Together they imply $\frac{\partial \mathcal{L}_J}{\partial \omega^\alpha} = 0$. This is a statement of the invariance of the action with respect to global group transformations. In Barbour's language \cite{barbour:scale_inv_particles}, this is the \emph{consistency requirement} that ensures that the linear momentum constraint is propagated by the equations of motion. Evaluating the partial derivatives we find
\equa{
    \diby{V}{\omega^\alpha} = \sum_I m_I \dbqii{I}{i} \lf(\frac{\taii{\alpha}{i}{j} + \taii{\alpha}{j}{i}}{2}\rt)\dbqii{j}{I}.
}

We interpret this equation of motion as a consistency condition on the form of allowed potentials. Potentials must be chosen such that this equation is satisfied. We note that this is philosophically different from standard gauge theory where the invariance of the action under global gauge transformations is an assumption rather than a result. In the case of rotations and translations the RHS is just zero so the consistency condition tells us that the potential must be independent of any global translations and rotations that we can perform on the system. This can be guaranteed by requiring that the potential be a function only of the distances $r_{IJ}$ between particles. This is in perfect agreement with our expectations of a Machian theory.

In the case of the scale transformations, the consistency condition takes the form
\equa{
    \sum_I \diby{V}{\bqii{i}{I}} \bqii{i}{I} = -2\sum_I m_I \dbqii{I}{i}\dbqii{i}{I}.
}
Using Euler's theorem, this condition tells us that the potential must be homogeneous of degree -2 in $\bar{q}$. This places a significantly more strict condition on the potential then the other symmetry transformations. For this reason, despite the interesting possibility discussed in \cite{barbour:scale_inv_particles}, it is still unclear whether this scale invariant model can really reproduce the physics of the real world. The apparent lack of scale invariance of the world is one of the most curious challenges of the Machian program.

\subsection{Summary of Results}

We will now collect our results regarding the CCM and how we used it to create a spatially relational theory of Newtonian particle dynamics. We started with our action, which in the case of the BB model was the Jacobi action, and substituted $\bar{q}$'s for $q$'s. We then did a free endpoint variation with respect to the auxiliary fields $\omega^\alpha$ and found it implied a linear constraint on the generalized momenta of the system as well as a consistency condition on the potential\footnote{Note that the dynamics of \emph{subsystems} is essentially constraint free.}. The consistency condition guarantees that the action be invariant under the global symmetries in question. The linear momentum constraint guarantees that the configuration of the system will have no unphysical motion relative to an absolute frame of rest. Thus, we saw explicitly how the CCM successfully eliminated the theory's dependence on the nonphysical symmetries present in the original theory. In our case, we were left with a relational theory of nonrelativistic particles. The remaining dynamics was given by varying the $q$'s and performing a gauge transformation to bring the equations of motions in standard Newtonian form.

In the next section, we will apply the CCM to a complex scalar field theory in Minkowski space. Specifically, we will use this method to eliminate the U(1) symmetry of the theory. Since we still have an absolute background Minkowski spacetime, we will not be left with a perfectly relational field theory. However, just like in the particle model, we will be successful at eliminating the dependence of the theory on an unphysical symmetry of the system. We will then be in a position to compare the CCM to Yang-Mills gauge theory.

\section{A U(1) Scalar Field Theory} \label{sec:U1sf}

Consider the complex scalar field theory living in Minkowski spacetime defined by the action (with ``mostly +'' signature for the metric):
\equa{\label{eq:phiaction}
    S_\phi = \int d^4x \, \lf[ m^2 \phi^\dag \phi - \partial_\mu \phi^\dag \partial^\mu \phi \rt].
}
This field theory enjoys a global U(1) symmetry owing to the invariance of the action under the transformation $\phi \rightarrow e^{i\gamma}\phi$ for some arbitrary constant $\gamma$. This theory is analogous to Newtonian mechanics on an absolute background where the absolute configuration space is no longer $\mathbb{R}^{3N}$ but rather a U(1) bundle over the collection of all possible field configurations in Minkowski spacetime. Performing a variation with respect to $\phi^\dag$ gives the Klein-Gordon equation for $\phi$
\equa{
    \Box\phi + m^2\phi = 0.
}
A similar Klein-Gordon equation is obtained for the complex conjugate of $\phi$ by varying $\phi$. Thus, the theory gives two Klein-Gordon equations for $\phi$ and $\phi^\dag$ coupled only by the complex conjugacy relations between $\phi$ and $\phi^\dag$. The global U(1) symmetry of the action implies a conserved Noether current
\equa{\label{eq:current}
    j_\mu = - i\lf( \partial_\mu\phi^\dag\,\phi -  \phi^\dag\partial_\mu \phi \rt)
}
obeying $\partial^\mu j_\mu = 0$.

In analogy to the particle model, we expect that the CCM will give us a linear momentum constraint that will set some generalized momentum, which would normally just be conserved, equal to zero. Intuitively, we can guess that this will be the vanishing of the Noether current. We will see in what follows that this is exactly what we obtain. Furthermore, we expect a consistency condition for the potential of the theory. Since our action is already globally U(1) invariant we would expect this consistency condition to be satisfied automatically. If our program is successful, we should expect to be left with a scalar field whose U(1) symmetry has been gauged away. This should just be a Klein-Gordon equation for a real scalar field.

\subsection{The Corrected Coordinate Method in Scalar Field Theory}

We implement the general formulation of the CCM by replacing $\phi$'s with $\bar{\phi}$'s in the action of \eq{phiaction}. In this case, the $\bphi$'s are given by
\equa{
    \bphi = e^{i\omega(\vec{x},t)} \phi;
}
which, when inserted into the action
\equa{
    S_{\bphi} = \int d^4x \, \lf[ m^2 \bphi^\dag \bphi - \partial_\mu \bphi^\dag \partial^\mu \bphi \, \rt]
}
reduces to
\equa{\label{eq:bphiaction}
    S_{\bphi} = \int d^4x \, \lf[ (m^2 -(\partial^\mu \omega)^2) \phi^\dag \phi - \partial_\mu \phi^\dag \partial^\mu \phi  + \partial^\mu j_\mu \rt].
}
$j_\mu$ is the Noether current given in \eq{current}. This is \emph{exactly} the action obtained by making the substitution $\partial_\mu \rightarrow D_\mu \equiv \partial_\mu + i\partial_\mu \omega$.

In this simple Abelian case, the CCM is equivalent to promoting the global U(1) symmetry to a local U(1) symmetry using a covariant derivative with a connection $A_\mu$ given explicitly by \equa{A_\mu = \partial_\mu \omega.} However, it is important to note that the connection here is \emph{not} the most general U(1) valued 1-form but, actually, it \emph{must} be the gradient of a function. In this sense, it is the ``pure gauge'' part of the Maxwell field. We will see that this will also be true in the non-Abelian case: the auxiliary fields can be mapped only to the ``pure gauge'' part of a Yang-Mills field and nothing more. In the Abelian case, it is easy to see that the curl of $A_\mu$ must be zero meaning that the curvature will be zero.

We can proceed now by working out the equations of motion of this theory and compare them to our expectations from the previous section. First, we note that the action of \eq{bphiaction} is manifestly invariant under variations of $\omega$. Thus, $\frac{\delta S}{\delta\omega} = 0$ exactly. The consistency condition is then automatically satisfied. Next, it is a short calculation to work out the variations with respect to the $\phi^\dag$ and $\phi$ fields. These give respectively
\begin{align} \label{eq:phieq}
    m^2 \phi + D_\mu D^\mu \phi &=0\text{, and} \\
    m^2 \phi^{\dag} + D_\mu^{\dag} D^{\dag\mu} \phi^{\dag} &=0. \label{eq:phieq2}
\end{align}

Finally, we can work out the Mach condition $\frac{\delta S}{\delta(\partial_\mu\omega)} = 0$. It implies
\equa{\label{eq:abelian_mach_cons}
    j^\mu - 2\phi^{\dag}\phi\, \partial^\mu\omega = 0.
}
In this case, the Mach condition is local because the auxiliary fields are functions of Minkowski space and not just of an arbitrary parameter $\lambda$. Writing $\phi$ and $\phi^\dag$ in terms of the real fields $\psi(\vec{x},t)$ and $\theta(\vec{x},t)$
\equa{
    \phi = \psi e^{i\theta}\text{ and}\qquad \phi^\dag = \psi e^{-i\theta};
}
we can rewrite the Mach condition as
\equa{\label{eq:mach_new}
    \partial_\mu \lf( \theta + \omega \rt) = 0.
}
This means that, aside from a constant, $\theta$ is just $-\omega$. Because the $\omega$ can be arbitrarily redefined using a banal transformation\footnote{See \cite{barbour:scale_inv_particles, sg:mach_time, sg:dirac_algebra} for more details on banal transformations.}, this means that the $\theta$ are arbitrary provided they satisfy the boundary conditions. Specifically, in the distinguished gauge, $\partial_\mu \theta = 0$, which is equivalent to $j^\mu = 0$: the vanishing of the Noether current. This is precisely what we would have expected for the U(1) invariant theory based on the analogy to the particle dynamics model.

Our last task for this section is to work out the equations for the amplitude $\psi$ of $\phi$ and $\phi^\dag$. This can easily be achieved by combining \eq{phieq} and \eq{phieq2} with the Mach condition of \eq{mach_new}. In both cases, we find
\equa{
    m^2 \psi + \Box \psi = 0.
}
Hence, we have a single Klein-Gordon equation for the real amplitude $\psi$ of $\phi$. We have exactly reproduced our expectations of the previous section.

\subsection{Summary of Results for U(1) Theory}

We showed that the CCM applied to the complex, U(1) invariant scalar field theory defined by the action of \eq{phiaction} gives a Klein-Gordon theory for the amplitude of the complex scalar fields where the phase is arbitrary provided the boundary conditions are satisfied. We succeeded in obtaining a theory that does not depend on the phase of the complex scalars and is thus indifferent to the original U(1) invariance. This is a kind of generalization of Relationalism applied to internal symmetries. We also found that the auxiliary fields introduced by the CCM can be mapped to the ``pure gauge'' part of a Maxwell connection. In the next section, we will see how these results generalize to non-Abelian connections giving a general relationship between the CCM and non-Abelian gauge theory.

\section{SU(N) Invariant Scalar Field Theory}\label{sec:SUNsf}

In this section we consider the SU(N) invariant scalar field theory with fundamental matter defined by the action
\equa{\label{eq:ymaction}
    S_{\phi_I} = \int d^4 x \, \lf[ m^2 \phi_I^\dag \phi^I - \partial_\mu \phi_I^\dag \partial^\mu \phi^I \rt]
}
where repeated indices are summed over and $I$ goes from 1 to $N$. To simplify notation, we will use vector notation for vectors in the internal space. Row vectors will be written in boldface capital Greek letters and column vectors will be indicated with a $\dag$ (when appropriate, the $\dag$ also denotes Hermitian conjugation). Matrices in the internal space will be specified by boldface capital roman characters. With this notation, $S_{\phi_I}$ takes the form
\equa{\label{eq:ymaction2}
    S_{\vphi} = \int d^4 x \, \lf[ m^2\vphid\vphi - \partial_\mu \vphid \partial^\mu \vphi \rt].
}

We can implement the CCM by making the replacement $\vphi\rightarrow\vG\vphi$ where $\vG \, \epsilon \, \text{SU(N)}$. Using the fact that $\vGd\vG = \mathbf{1}$ and the resulting identity $(\partial_\mu \vGd) \vG + \vGd\partial_\mu \vG = 0$, the action in terms of the barred $\vphi$ reduces to:
\begin{multline}\label{eq:ymbaction}
    S_{\mathbf{\bar{\Phi}}} = \int d^4 x \, \lf[ \vphid \lf( m^2\mathbf{1} - \partial_\mu\vGd \partial^\mu\vG \rt) \vphi - \partial_\mu \vphid \partial^\mu \vphi \right. \\
    \left. - \lf( \partial_\mu\vphid ( \vGd \partial^\mu\vG ) \vphi - \vphid ( \vGd\partial_\mu\vG ) \partial^\mu\vphi \rt) \rt].
\end{multline}
Comparing this to the action $S_{A}$ one obtains by promoting the global SU(N) symmetry to a local symmetry by substituting partial derivatives for covariant derivatives of the form $\mathbf{D}_\mu = \mathbf{1}\partial_\mu + \vA_\mu$
\equa{
    S_{A} = \int d^4 x \, \lf[ \vphid \lf( m^2\mathbf{1} - \vAd_\mu \vA^\mu \rt) \vphi - \partial_\mu \vphid \partial^\mu \vphi - \lf( \partial^\mu\vphid \vAd_\mu \vphi - \vphid \vA_\mu \partial^\mu\vphi \rt) \rt]
}
we find that $S_{\mathbf{\bar{\Phi}}} = S_{A}$ provided
\equa{\label{eq:map}
    \vA^\mu = \vGd \partial^\mu\vG.
}

This is the key relation that allows us to map the auxiliary field of the CCM to a non-Abelian connection. In differential form notation, this relation is written $\vA = \vGd d\vG$. The curvature, $\mathbf{F} = d\vA + \vA \wedge \vA$ can be calculated by making use of the related identities $\vGd\vG = \mathbf{1}$, $(d\vGd) \vG = - \vGd d\vG$, and $d\vGd \wedge d\vG = 0$. It is easily seen to vanish. Thus, the auxiliary fields are mapped to the flat part of the connection only. These are the ``pure gauge'' degrees of freedom as they contribute nothing to the Yang-Mills self-coupling term $\star \mathbf{F} \wedge \mathbf{F}$ which involves only the curvature. The fact that the auxiliary fields of best-matching represent only the ``pure gauge'' degrees of freedom of the Yang-Mills connection agrees with our intuition since the auxiliary fields are meant to shift the configuration of the system along its gauge orbits and should not affect the local dynamics of the system other then by imposing global constraints on the total momentum. This, however, raises an interesting question: what does it mean physically to generalize the CCM to arbitrary Yang-Mills connections? This will introduce new \emph{physical} degrees of freedom into the system that are normally interpreted as gauge bosons. Is there a \emph{physical} motivation for introducing these gauge bosons in light of a Machian program? These questions will be revisited in \scn{generalizations}.

\subsection{General Derivation}\label{sec:gen_der}

It is worthwhile at this point to outline a more general derivation of \eq{map}. In order for the CCM to agree with the general gauging method rather than the gauging method applied to specific potentials, we need an equivalence between the two quantities $\partial_\mu \mathbf{\bar{\Phi}}$ and $\mathbf{D}_\mu \vphi$. This is nearly achieved exactly but needs some tweaking. The nature of this tweaking will lead us to a modified formulation of the CCM. Instead of having an identification of $\partial_\mu \mathbf{\bar{\Phi}}$ with $\mathbf{D}_\mu \vphi$ we have the following easily verified relation
\equa{
    \mathbf{G}^{-1} \partial_\mu \mathbf{\bar{\Phi}} = \mathbf{D}_\mu \vphi,
}
where $\vA = \mathbf{G}^{-1} \partial_\mu \vG$. In the specific case $\vG \, \epsilon \, \text{SU(N)}$ we have $\mathbf{G}^{-1}=\vGd$. Using this we can not only prove that $\vA$ is identical to the expression in \eq{map} but we can also show that the kinetic terms of the CCM and the gauged theory are equivalent. That is,
\equa{
    \mathbf{D}_\mu \vphid \mathbf{D}^\mu \vphi = \partial_\mu \mathbf{\bar{\Phi}}^\dag \partial^\mu \mathbf{\bar{\Phi}}.
}

We note two important points. The first is that there wouldn't necessarily be a correspondence between standard gauge theory and the CCM if the gauge group wasn't such that $\mathbf{G}^{-1}=\vGd$. For example, in the particle model, this is true for the translations and the rotations but not for the scale transformations\footnote{This makes the correspondence between gauge theory and the CCM less obvious for the scale transformations. See \scn{mod_ccm} for an alternative to the CCM which makes the general correspondance more clear.}. The second interesting observation is that, in the theory of \eq{ymaction2}, $\mathbf{\Pi}_\mu = \partial_\mu \mathbf{\Phi}$ is the momentum conjugate to $\mathbf{\Phi}$. Thus, the correspondence to gauge theory is exact if we apply the transformation $(\vphi,\mathbf{\Pi})\rightarrow(\vG\vphi,\vG^{-1}\mathbf{\Pi})$. This transformation preserves the Poisson brackets between $\vphi$ and $\mathbf{\Pi}$ and, thus, is canonical as far as these variables are concerned. Though the Poisson brackets between these variables and the auxiliary fields and their momenta will be modified, the auxiliary fields are arbitrary so, at least intuitively, we would expect that the local physics should remain unchanged. This possibility suggests an alternative to the original CCM which will be explored in more detail in \scn{mod_ccm}.

\subsection{Equations of Motion}
\label{sec:ymeom}

In this section we will compute the equations of motion given by the CCM. Our motivation is twofold. First, we seek to compare these results with the results of the Abelian theory to allow us to gain some intuition for their meaning. Second, we would like to show that the free endpoint variation of the auxiliary fields is equivalent to the standard variation of a flat gauge connection. Only after accomplishing this can we claim that the CCM is equivalent to flat non-Abelian gauge theory.

Before performing the variations it will be constructive to rewrite the action of \eq{ymbaction} in a more familiar form. We first note that the SU(N) invariance of the unaltered action of \eq{ymaction} implies that there will be a conserved Noether current of the form
\equa{\label{eq:ymcurrent}
    j^\mu_\alpha = \vphid \vt_\alpha \partial^\mu \vphi - \partial^\mu \vphid \vt_\alpha \vphi,
}
which is a direct generalization of the U(1) current. Using the fact that $\vG = \exp(\omega^\alpha \vt_\alpha)$ we can differentiate to find
\equa{\label{eq:dG}
    \partial_\mu \vG = \partial_\mu \omega^\alpha \vG \vt_\alpha.
}
Incidently, this leads to a more direct way of writing the connection, $\vA_\mu$, in terms of the auxiliary fields $\omega^\alpha$: $\vA_\mu = \partial_\mu \omega^\alpha \vt_\alpha$. Or, in terms of the components $A_\mu^\alpha$ such that $\vA_\mu = A_\mu^\alpha \vt_\alpha$: $A_\mu^\alpha = \partial_\mu \omega^\alpha$ which is the natural generalization of the U(1) result. With the help of \eq{dG} and the current of \eq{ymcurrent} we can rewrite the action of \eq{ymbaction} as
\equa{
    S_{\mathbf{\bar{\Phi}}} = \int d^4 x \, \lf[ \vphid \lf( m^2\mathbf{1} - \partial_\mu \omega^\alpha\vtd_\alpha \partial^\mu\omega^\beta\vt_\beta \rt) \vphi - \partial_\mu \vphid \partial^\mu \vphi - \partial_\mu\omega^\alpha j^\mu_\alpha \rt]
}
which is the analogue of \eq{bphiaction}.

The action is now in a form where it is trivial to compute the variations. We will start by proving that the normal variation of $\vA_\mu$ is equivalent to the free endpoint variation of the auxiliary fields. The consistency condition is $\frac{\delta S_{\mathbf{\bar{\Phi}}}}{\delta\omega^\alpha}=0$. This is automatically satisfied since $S_{\mathbf{\bar{\Phi}}}$ does not explicitly depend on $\omega^\alpha$. In the non-Abelian case, we have only the Euler-Lagrange equations
\equa{
    \frac{\delta S_{\mathbf{\bar{\Phi}}}}{\delta A_\mu} - \partial_\nu\lf( \frac{\delta S_{\mathbf{\bar{\Phi}}}}{\delta (\partial_\nu A_\mu)} \rt) = 0,
}
where $A_\mu = \partial_\mu \omega$. But, by looking at the action we can clearly see that $\frac{\delta S_{\mathbf{\bar{\Phi}}}}{\delta(\partial_\mu A_\nu)}=0$ and the Euler-Lagrange equations reduce simply to the Mach condition $\frac{\delta S_{\mathbf{\bar{\Phi}}}}{\delta A_\mu}=\frac{\delta S_{\mathbf{\bar{\Phi}}}}{\delta (\partial_\mu\omega^\alpha)}= 0$. We note that this happens because of two contributing facts: 1) the consistency condition is automatically satisfied, and 2) $\partial_\mu \omega$ plays the role of a momentum in the CCM while $A_\mu$ plays the role of a configuration space variable in the standard gauge theory. This swapping of roles is crucial to the equivalence of both variations.

The Mach condition can now easily be computed. It is
\equa{\label{eq:non_a_mach}
    j^\mu_\alpha = -\partial^\mu \omega^\beta \, \vphid \lf( \vt_\alpha \vt_\beta + \vt_\beta \vt_\alpha \rt) \vphi.
}
In the fundamental representation of SU(2), the $\vt$'s are just the Pauli matrices and the above gives $j^\mu_\alpha=-2i\partial^\mu \omega^\alpha \vphid\vphi$. In this form, it is clearly a direct generalization of the Mach condition (\ref{eq:abelian_mach_cons}) from the Abelian case. To get a handle on the general non-Abelian case, we rewrite the fields $\vphi$ explicitly in terms of the norm $\psi = \sqrt{\vphid\vphi}$ and the quantity $\vphig$ such that
\equa{
   \vphi = \psi \vphig.
}
In these variables, it is clear that the Mach condition (\ref{eq:non_a_mach}) does not depend on the norm $\psi$ but only on $\vphig$. If we insert $\psi$ and $\vphig$ into (\ref{eq:non_a_mach}) and use (\ref{eq:ymcurrent}) we find that the factors of $\psi$ cancel leaving us with
\equa{
    \partial^\mu\vphigd \vt_\alpha \vphig - \vphigd \vt_\alpha \partial^\mu\vphig = \vphigd \lf( \vt_\alpha \vt_\beta + \vt_\beta\vt_\alpha \rt) \vphig \partial^\mu \omega^\beta.
}
This equation may not be as enlightening as the U(1) or the SU(2) case but it will be very useful in simplifying the equations of motion for $\vphi$ and $\vphid$. Note that, in the distinguished representation, $\partial_\mu \omega^\beta = 0$ and the RHS is zero. This implies that the Noether current should vanish in agreement with our expectations.

Our last task is to work out the equation of motion for the scalar fields. Variations with respect to $\vphid$ and $\vphi$ give, respectively,
\begin{align}
    m^2 \vphi + \mathbf{D}_\mu \mathbf{D}^\mu \vphi &=0\text{, and} \label{eq:eom_non_a1}\\
    m^2 \vphid + \mathbf{D}_\mu^{\dag} \mathbf{D}^{\dag\mu} \vphid &=0.\label{eq:eom_non_a2}
\end{align}
We can now carefully rewrite these equations of motion in terms of $\psi$ and $\vphig$. We can then multiply (\ref{eq:eom_non_a1}) on the left by $\vphigd$ and add to it the product of (\ref{eq:eom_non_a2}) with $\vphig$. Using the property $\vphigd\vphig = 1$, which is just a result of the definition of $\vphig$, and the resulting identities
\begin{align}
    \partial^\mu\vphigd \vphig + \vphigd \partial^\mu\vphig &=0, \qand \\
    \Box\vphigd \vphig + \vphigd \Box\vphig &=0,
\end{align}
we find that the equations of motion can be combined to give
\equa{
    (\Box + m^2)\psi = \lf(\partial^\mu\vphigd \vt_\alpha \vphig - \vphigd \vt_\alpha \partial^\mu\vphig - \vphigd \lf( \vt_\alpha \vt_\beta + \vt_\beta\vt_\alpha \rt) \vphig \partial^\mu \omega^\beta\rt)\psi.
}
Of course, the RHS is just proportional to the Mach condition. Thus, the equations of motion of the complete system reduce simply to
\equa{
    m^2\psi + \Box\psi =0,
}
which is the Klein-Gordon equation in terms of $\psi$ only, plus the Mach condition, which is in terms of $\vphig$ and $\vphigd$ only. This is completely analogous to the Abelian case.

\subsection{Summary of Results}

We have shown that the CCM applied to a particular symmetry is equivalent to the standard gauging of that symmetry using a flat connection of the form $\vA = \vG^{-1} d\vG$ if $\vG^{-1} = \vGd$. Then, we worked out the equations of motion of a massive SU(N) invariant scalar field theory with fundamental matter using the CCM. We found that this system reduced to a Klein-Gordon equation for the norm $\psi$ of the scalar fields and a Mach condition in terms of $\vphig$ and $\vphigd$. This is a clear generalization of the Abelian case where a general notion of Relationalism was realized through the fact that the real physical theory was in terms of $\psi$ and was decoupled from internal degrees of freedom.

\section{Modified Corrected Coordinate Method} \label{sec:mod_ccm}

In \scn{cc_particle}, we noticed that the CCM is a least-squares minimization that minimizes the square of the quantity:
\equa{\label{eq:min2}
    G (\dot q + \dot{\omega}^\alpha t_\alpha q).
}
This is achieved by inserting the corrected coordinates $\bar{q} = Gq$ into the original action which contains the derivatives $\dot{\bar{q}} = G (\dot q + \dot{\omega}^\alpha t_\alpha q)$ in a quadratic kinetic term. The vanishing of the variation of the action will guarantee that the length of $\dot{\bar{q}}$ is minimized. However, in \scn{gen_der}, we showed that for the correspondence between gauge theory and the CCM to be exact we needed a correspondence of the form:
\equa{
    G^{-1}d(Gq) = Dq,
}
where $d$ denotes the exterior derivative (in \eq{min2} this would just represent a ``dot'') and $D$ is the covariant derivative of a connection $A$. In this case, there is an equivalence when $A = G^{-1}dG$. Thus, in standard gauge theory with a flat connection, the quantity being minimized is the length of
\begin{align}
    G^{-1}d(Gq) &= dq + (dG) q \notag \\
                &= \dot{q} + \dot{\omega}^\alpha t_\alpha q. \label{eq:min3}
\end{align}
Comparing \eq{min2} with \eq{min3}, we see that the difference between gauge theory with a flat connection and the original CCM is a factor of $G^{-1}$ in front of the quantity to be minimized.

We can ask ourselves about the physical significance of this factor of $G^{-1}$. From the point of view of the $\chi^2$ minimization, this is just like taking the whole system in both initial and final configurations and translating it backwards along the group orbit so that the original configuration is aligned with the identity element. It's like redefining the zero of the auxiliary fields $\omega^\alpha$. Thus, based on physical arguments, it is clear that the extra factor of $G^{-1}$ should not change the final physical theory. This is not so obvious from the point of view of the mathematics. We refer the reader to \cite{sg:dirac_algebra} for a detailed proof of the mathematical equivalence between the two methods which relies on gauge invariance.

Inspired by this physical argument and the correspondence between gauge theory and best-matching, we suggest a new prescription for implementing Mach's principle. Instead of making the substitution
\equa{
    q \rightarrow \bar{q} = Gq,
}
which implies
\begin{equation}
    \dot{q} \rightarrow \dot{\bar{q}} = G (\dot q + \dot{\omega}^\alpha t_\alpha q),
\end{equation}
and inserting these into the original action of the theory, we suggest the new substitutions
\begin{align}
    q &\rightarrow \bar{q} = Gq \\
    \dot{q} &\rightarrow G^{-1}\dot{\bar{q}} = (\dot q + \dot{\omega}^\alpha t_\alpha q).
\end{align}
This \emph{modified} CCM is completely equivalent to gauge theory with a flat connection. As is outlined in \cite{sg:dirac_algebra}, the Dirac algebra of the modified theory is much easier to work with. This combined with the equivalence to gauge theory leads us to believe that the modified method is more natural.

\subsection{Comparison to the Original Formulation}

We now ask ourselves what the modified CCM looks like in the original formulation of the CCM. In this formulation the corrected coordinates are given by
\equa{
     \bqii{i}{I} = q^{i}_{I} + \epsilon\lf( t^{i} + \epsilon^{ijk} r^{j} q^{k}_I + d q^i_I \rt)
}
and the derivatives are
\equa{
        \dot{\bar q}^i_I = \dot{q}^{i}_{I} + \epsilon\lf( \dot{t}^{i} + \epsilon^{ijk} \dot{r}^{j} q^{k}_I + \dot{d} q^i_I \rt)+ \epsilon\lf( \epsilon^{ijk} r^{j} \dot{q}^{k}_I + d \dot{q}^i_I \rt)
}
where the third term is hard to deal with but doesn't seem to have an effect on the physical theory (this has been discussed already in \scn{org_form}). In the modified approach, we keep the same corrected coordinates $\bqii{i}{I}$ but substitute
\equa{
    \dot{q} \rightarrow G^{-1}\dot{\bar{q}} = (\dot q + \dot{\omega}^\alpha t_\alpha q).
}

In the limit of infinitesimal $\omega^\alpha$ we can use the substitutions of (\ref{eq:new_2_old}) to find that, using the original variables, the modified CCM makes the substitution
\equa{
    \dot{q}^i_I \rightarrow \dot{q}^{i}_{I} + \epsilon\lf( \dot{t}^{i} + \epsilon^{ijk} \dot{r}^{j} q^{k}_I + \dot{d} q^i_I \rt).
}
Thus, the modified CCM is just the original CCM but without the awkward term. The disappearance of this term in the new formulation is the analogue of the simplifications that occur in the Dirac algebra. This is yet another argument for the naturalness of the modified CCM.

\section{Curved Generalizations}
\label{sec:generalizations}

Recall that the CCM was a technique derived to implement Mach's principle by matching different configurations of a system using a least-squares type minimization. As such, it implied a very specific form for the $\lambda$-dependent gauge connection on configuration space. In this section, we consider the possibility of making a different choice of gauge connection from that implied by the CCM. In particular, we will consider a Yang-Mills connection. This will necessarily introduce new degrees of freedom on top of the auxiliary fields corresponding to the part of the gauge connection contributing to nonzero curvature. That is why we can treat this as a generalization of the CCM. More work, however, will be required to precisely determine the relationship between Mach's principle and the curved part of the connection. 

We will study the simplest case of a U(1) invariant field theory where the new degrees of freedom are seen to couple to the physical degrees of freedom of the scalar fields. Because of this coupling, these new fields are either trivial, in the case of no Yang-Mills self-interaction, or dynamical, in the case of a nonzero self-interaction term where they are equivalent to the Maxwell field.

\subsection{Scalar U(1) Field Theory With Curvature (No Yang-Mills Terms)}

Consider the action
\equa{
    S_{\phi, A_\mu} = \int d^4 x \, \lf[ m^2 \phid\phi - D_\mu \phid D^\mu\phi \rt]
}
where $D_\mu = \partial_\mu + iA_\mu$. Now, however, we will allow $A_\mu$ to have two pieces: one that corresponds to its flat part which is the gradient, $\partial_\mu\omega$, of the auxiliary fields and another that corresponds to curved part which we will denote simply as $A'_\mu$. Thus,
\equa{\label{eq:A}
    A_\mu = A'_\mu + \pmw{\mu}.
}
The action can now be written
\equa{
    S_{\phi, A_\mu} = \int d^4 x \, \lf[ (m^2 - A^2) \phid\phi - \partial_\mu \phid \partial^\mu\phi + iA_\mu j^\mu \rt].
}
$j^\mu$ is the usual U(1) current given by \eq{current}.

A variation with respect to the $\phid$'s and the $\phi$'s gives, just as before
\begin{align}
    m^2 \phi + D_\mu D^\mu \phi &=0\text{, and} \label{eq:phiA1}\\
    m^2 \phi^{\dag} + D_\mu^{\dag} D^{\dag\mu} \phi^{\dag} &=0. \label{eq:phiA2}
\end{align}
Now, however, $D_\mu$ is the more general covariant derivative defined by a general connection $A_\mu$ of \eq{A}. At first sight, this would seem to give drastically new physics, however, with no Yang-Mills self-interaction term, the dynamics will force $A'_\mu = 0$ giving the same results as before.

To see how this happens, we perform the variation of the action with respect to $A_\mu$. Keeping our results from \scn{ymeom} in mind we know that this variation will be equivalent, as far as the flat piece is concerned, to the free endpoint variation. The equations of motion that one obtains are:
\equa{\label{eq:Avar}
    A_\mu \phid\phi - \frac{i}{2}j_\mu = 0.
}
Parameterizing $\phi$ as before in terms of the norm $\psi$ and the phase $\theta$ such that $\phi = \psi e^{i\theta}$ we find that this equation takes the form
\equa{
    \partial_\mu( \theta + \omega ) = A'_\mu.
}
But, since $A'_\mu$, by definition, cannot be the gradient of anything, it must be zero.

\subsection{Scalar U(1) Field Theory With Curvature (With Yang-Mills Terms)}

In order to obtain a nontrivial theory for $A'_\mu$ we must add a self-interaction term to the action. The simplest and most standard term to add would be the Yang-Mills interaction term $\star F \wedge F$ where $F$ is the curvature 2-form of $A$. The new term in the action is
\equa{
    S_{YM} = \int d^4 x \, \frac{1}{2} F_{\mu\nu}(A')F^{\mu\nu}(A').
}

The only variation that is changed by adding this term is the $A$ variation. The effect is to supplement \eq{Avar} with a term proportional to the gradient of $F$. The resulting equation of motion is
\equa{\label{eq:maxwell}
    A_\mu \phid\phi - \frac{i}{2} j_\mu = -\partial^\nu F_{\mu\nu}.
}
Using the standard current $J_\mu$ for a U(1) scalar field
\equa{
    J_\mu = \frac{i}{2} j_\mu - A_\mu \phid\phi
}
\eq{maxwell}, which is the analogue of the Mach condition, is just the standard Maxwell equation
\equa{
    J_\mu = \partial^\nu F_{\mu\nu}.
}
We obtain, as we expected, a Maxwell field coupled to a complex scalar field.

We can write \eq{maxwell} in terms of $A'_\mu$, $\omega$, $\psi$, and $\theta$ in order to connect with our work on the CCM. In these variables, \eq{maxwell} becomes
\equa{
    (A'_\mu + \partial_\mu(\omega + \theta))\psi^2 = -\partial^\nu F_{\mu\nu}(A').
}
Because the RHS depends only on $A'_\mu$, this can only be satisfied for arbitrary $A'_\mu$ and nontrivial $\psi$ provided
\begin{align}
    \partial_\mu(\omega + \theta) & = 0\, \text{, and} \label{eq:wt} \\
    A'_\mu \psi^2 &= -\partial^\nu F_{\mu\nu}(A'). \label{eq:Apsi}
\end{align}
Note that the Maxwell equations couple the physical degrees of freedom, $A'$ and $\psi$, to each other only through \eq{Apsi} and the gauge degrees of freedom, $\omega$ and $\theta$, to each other only through \eq{wt}. As a result, the phase of the scalar field is still pure gauge while the dynamics of the norm $\psi$ will be physically altered.

To see how the dynamics of $\psi$ will be altered, we can work out the equations of motion (\ref{eq:phiA1}) and (\ref{eq:phiA2}) in terms of the full connection $A_\mu = A'_\mu + \partial_\mu \omega$. Note that we can take the divergence of (\ref{eq:Apsi}) to get
\equa{
    \partial_\mu A'^\mu + 2 A'^\mu \partial_\mu \psi = 0.
}
Using this and \eq{wt}, we can rewrite (\ref{eq:phiA1}) and (\ref{eq:phiA2}). In both cases, they take the form
\equa{
    \lf( m^2 - A'^2 + \Box \rt) \psi = 0.
}
Thus, we again have a Klein-Gordon theory in terms of the norm $\psi$. Now, however, the mass term is shifted by the square of the curved part of the Maxwell field.

\subsection{Summary of Results}

We found that the only way to get a nontrivial theory after allowing for curvature in the scalar U(1) invariant theory is to add a self-interaction term to the action. When we add a Yang-Mills term, we obtained a Maxwell theory for the curved part of the connection $A'$ coupled to a Klein-Gordon theory for the norm of the complex scalar field. The nonphysical phases decouple from the physical sector of the theory leaving them arbitrary. This seems to indicate that the theory is still, in a generalized sense, Relational.

\section{Discussion / Outlook}

In this work we have achieved two main goals:

\begin{enumerate}
    \item We have introduced a new formulation of the CCM, inspired by group theory, which can deal with large values of the auxiliary fields. This formulation writes the corrected coordinates in the form:
          \equa{\bar{q}^i_I = G(\omega^\alpha)^i_j q^j_I = \exp(\omega^\alpha \taii{\alpha}{i}{j}) q^j_I}
        and uses the group generators $\taii{\alpha}{i}{j}$ explicitly. This method presents several advantages over the pervious method \cite{barbour:scale_inv_particles} including: 1) we can now deal with the $q$ dependence of group elements exactly so we have a better handle on the kinetic term, 2) we can consider the action of arbitrary Lie groups which allows us to make general claims about the constraints implied by these symmetries\footnote{See \cite{sg:dirac_algebra} for further details on the constraints of general symmetry groups.}, and 3) the mathematics are more straightforward and are free of awkward terms that exist in the original approach.
    \item We have established an isomorphism between the auxiliary fields used in best-matching (and the CCM) and the components of a flat connection used in gauge theories. This isomorphism, given by \eq{map}, has uncovered a deep connection between Mach's principle, or more precisely the requirement of Relationalism, and the gauge principle used in Yang-Mills gauge theories. Though we discovered this relationship by applying the CCM to Yang-Mills theories, we can turn this around and say that the CCM implements Mach's principle by using the techniques of gauge theory. This connection has brought to light several important observations.

        First, the free endpoint variation of the auxiliary fields is \emph{only} equivalent to the variation of a curvature free connection because the consistency conditions $\frac{\delta S}{\delta \omega} = 0$ are automatically satisfied. This is true because in standard gauge theory one always starts with an action that is invariant under global gauge transformations. However, from the point of view of best-matching, this is not necessary and the consistency conditions can be seen as constraints that must be satisfied by the nonkinetic terms of the theory. In this sense, global gauge invariance is seen as a requirement for the consistency of the equations of motion rather than an assumption of the theory. Best-matching is, thus, a more powerful framework than standard gauge theory in this regard.

        Second, in order for complete equivalence between the two approaches in the general context, the CCM needs to be modified slightly. Instead of writing
        \equa{
            dq \ra d(Gq) = (dG)q + G dq
        }
        we have
        \equa{
            dq \ra G^{-1}d(Gq) = dq + G^{-1}(dG)q = dq + (d\omega^\alpha)t_\alpha q,
        }
        where $d$ stands for the exterior derivative over the base manifold. Because, as we have shown with physical arguments, this modified version of the CCM is equivalent to the original formulation we are led to suggest it as a more natural implementation of best-matching. The main difference between this modified approach and the original CCM is the disappearance of awkward, nonphysical terms. This leads us to conclude that our modified CCM is the more nature and mathematically cleaner way to implement Mach's principle in best-matching.

        Finally, we note the valuable intuition gained by looking at the $U(1)$ and $SU(N)$ models. From these models it is clear that the CCM is implementing a generalized form of Relationalism where only the physical part of the field (the norms in this case) are actually evolving while the nonphysical phases become pure gauge degrees of freedom. The relationship with gauge theory suggests a natural generalization to this where new curvature degrees of freedom can be introduced by adding a Yang-Mills interaction term and by allowing the connection to have nonflat components. The physical meaning, however, of these generalizations from a relational point of view remains unclear. A final benefit of considering the connections to gauge theory is to import valuable intuition and sophisticated mathematical tools into the language of best-matching. Because best-matching can be used to derive geometrodynamics, this might provide valuable insight into the gauge theory nature of general relativity.

%    \item We discovered that the procedure of first best-matching the theory of Newtonian particle mechanics then quantizing commutes with the procedure of first quantizing then best-matching. That is, we saw that if we treat the Schrodinger theory of $N$ nonrelativistic particles as arising from an action principle then it is possible to apply the best-matching procedure to the invariance of this theory under diffeomorphims generated by the global Killing vectors of the configurations space. The theory obtained in this way was exactly the theory obtained by quantizing Barbour-Bertotti theory.

%        This example is particularly thought provoking because the way in which the corrected-coordinate method is implemented is nearly identical to the way in which it is implemented in Barbour and Bertotti's formulation of geometrodynamics. Though there are glaring differences between the two theories, it is somewhat surprising to note that General Relativity shares so many features with a quantized model of relational particles. It might be worth investigating just how far these particle models can be pushed to reproduce the essential features of gravity.
\end{enumerate}

% -------------------------------------- Acknowledgments ---------------------------------------
\begin{acknowledgments}
    I am eternally grateful to Julian Barbour for explaining, in the backdrop of the historic College Farm, the key ideas behind his beautiful approach to Mach's principle. Also, I am very thankful for refreshing discussions with Henrique Gomes and Hans Westman that have pushed me deeper into thoughts regarding the relationship between Mach's principle and gauge theory. Finally, I would like to thank Lee Smolin for his motivation and guidance. Research at the Perimeter Institute is supported in part by the Government of Canada through NSERC and by the Province of Ontario through MEDT. I also acknowledge support from an NSERC Postgraduate Scholarship, Mini-Grant MGA-08-008 from the Foundational Questions Institute (fqxi.org), and from the University of Waterloo.
\end{acknowledgments}

% ---------------------------------------- Appendices (optional) -------------------------------

%\appendix     % For many appendix sections
%\appendix*    % For a single appendix section.

%\section{my appendix}

% --------------------------------------- Bibliography ------------------------------------------
\bibliography{mach}
\bibliographystyle{utphys}

\end{document}